\documentclass[fleqn,10pt]{wlscirep}
\usepackage[utf8]{inputenc}
\usepackage[T1]{fontenc}
\usepackage{multirow}
\usepackage{tikz}
\usepackage{fontawesome5}
\usetikzlibrary{trees}

\title{SeizeIT2: Wearable Dataset Of Patients With Focal Epilepsy}

\author[1,*]{Miguel Bhagubai}
\author[1]{Christos Chatzichristos}
\author[2]{Lauren Swinnen}
\author[2]{Jaiver Macea}
\author[1]{Jingwei Zhang}
\author[4]{Lieven Lagae}
\author[4]{Katrien Jansen}
\author[5]{Andreas Schulze-Bonhage}
\author[6]{Francisco Sales}
\author[7]{Benno Mahler}
\author[8]{Yvonne Weber}
\author[2]{Wim Van Paesschen}
\author[1,3]{Maarten De Vos}

\affil[1]{Department of Electrical Engineering (ESAT), STADIUS Center for Dynamical Systems, Signal Processing and Data Analytics, KU Leuven, 3001 Leuven, Belgium}
\affil[2]{Laboratory for Epilepsy Research, UZ Leuven, 3000 Leuven, Belgium}
\affil[3]{Department of Development and Regeneration, KU Leuven, 3000 Leuven, Belgium}
\affil[4]{Department of Pediatric Neurology, UZ Leuven, 3000 Leuven, Belgium}
\affil[5]{Epilepsy Center, University Medical Center, Freiburg University, 79106 Freiburg, Germany}
\affil[6]{Epilepsy Reference Center, Coimbra University Hospital, 3004-504 Coimbra, Portugal}
\affil[7]{Department of Neurology, Karolinska University Hospital, 171 77 Stockholm, Sweden}
\affil[8]{Department of Epileptology and Neurology, RWTH University of Aachen, 52074 Aachen, Germany}
\affil[*]{corresponding author: Miguel Bhagubai (miguel.bhagubai@esat.kuleuven.be)}

\begin{abstract}
The increasing technological advancements towards miniaturized physiological measuring devices have enabled continuous monitoring of epileptic patients outside of specialized environments. The large amounts of data that can be recorded with such devices holds significant potential for developing automated seizure detection frameworks. In this work, we present SeizeIT2, the first open dataset of wearable data recorded in patients with focal epilepsy. The dataset comprises more than 11,000 hours of multimodal data, including behind-the-ear electroencephalography, electrocardiography, electromyography and movement (accelerometer and gyroscope) data. The dataset contains 886 focal seizures recorded from 125 patients across five different European Epileptic Monitoring Centers. We present a suggestive training/validation split to propel the development of AI methodologies for seizure detection, as well as two benchmark approaches and evaluation metrics. The dataset can be accessed on OpenNeuro and is stored in Brain Imaging Data Structure (BIDS) format.
\end{abstract}

\begin{document}

\tikzstyle{every node}=[draw=black,thick,anchor=west]
\tikzstyle{selected}=[draw=red,fill=red!30]
\tikzstyle{optional}=[dashed,fill=gray!50]
\tikzstyle{file}=[draw=none]

\flushbottom
\maketitle

\thispagestyle{empty}

\section*{Background \& Summary}

Epilepsy is one of the most common neurological disorders, affecting around 1\% of the global population \cite{jaiver1}, and seizures are its primary clinical manifestation \cite{jaiver2}. Seizures are transient phenomena resulting from the activation and distribution of abnormal or excessive neuronal activity within limited or diffuse brain networks \cite{jaiver3}. The signs observed by others and the symptoms reported by the patient depend on the brain region(s) where the abnormal activity arises and its further propagation pattern \cite{jaiver4}. Therefore, patients can present with different manifestations, including but not restricted to altered consciousness, abnormal visual, behavioral, auditory, sensorimotor, psychic, and autonomic phenomena, or a combination of these \cite{jaiver4, jaiver5}. The International League Against Epilepsy (ILAE) classifies seizures into three main groups based on their onset: focal, when the network is limited to one hemisphere; generalized, defined by the “rapid engaging of bilaterally distributed networks”; and unknown if the onset cannot be identified \cite{jaiver5}. The unpredictability of epileptic seizures has a negative impact on the quality of life in many patients \cite{jaiver6}, and epilepsy is associated with an increased risk for accidents and mortality \cite{jaiver7, jaiver8}.

After the diagnosis of epilepsy, patients receive pharmacological treatments for seizure control. In some drug-resistant cases, additional dietary or surgical treatments may be beneficial. In outpatient settings, documenting seizures is of great importance for monitoring the course of the disease and adjusting the treatments. Currently, clinicians rely on the seizures reported by the patient in diaries. However, seizure diaries have shown a documentation sensitivity of less than 50\% \cite{jaiver9, bteabs1}.

Furthermore, 30\% of patients with epilepsy have drug-resistant epilepsy\cite{jaiver11}. In many of these cases, a comprehensive evaluation is done, including recording and analysing seizures using video electroencephalography (vEEG). As vEEG is a costly and time-consuming examination, patient selection is crucial for effective use of resources. Due to the unpredictability of seizures, it is uncertain whether a seizure will be captured during a vEEG admission, reducing the diagnostic yield and delaying clinical decision-making \cite{jaiver12, jaiver13}. Therefore, clinicians require more efficient methods to improve patient monitoring. 

The development of small wearable recording devices has surged in recent years. These devices can measure physiological data for long periods outside the hospital, enabling continuous monitoring of patients \cite{wearables}. Wearable EEG technologies are often designed to discreetly encapsulate a reduced set of electrodes intended to measure atypical regions of the scalp, such as behind-the-ear. Devices like the Sensor Dot (SD), from Byteflies \cite{byteflies}, were developed for the purpose of monitoring epilepsy patients during their daily lives. The development of the SD device was accomplished within the SeizeIT1 study, where patients with focal epilepsy were monitored in-hospital with vEEG, together with an additional set of electrodes placed behind-the-ear, that mimicked the wearable setup and its data characteristics. Preliminary validation studies using this recording setup showed promising results for the detection of seizures in a controlled environment with a simulated wearable behind-the-ear EEG (bte-EEG) setup \cite{bteEEG, kaat1}. It was concluded that the performance between scalp and wearable EEG detection algorithms was similar in a small group of patients with temporal lobe seizures, proving the potential of using such a modality for seizure detection. Other types of wearables have been used to monitor different physiological signals, such as electrocardiography (ECG), electromyography (EMG), accelerometry (ACC), gyroscope data (GYR), electrodermal activity (EDA) and combinations of these \cite{multimodal}. When data is recorded outside of the hospital, clinicians do not have video information to assess behavioural changes caused by seizures, such as spasms, tonic or clonic episodes. By combining information from different physiological signals, using multimodal wearables, the detection of seizures can be improved \cite{multi_1, multi_2}.

The use of wearable devices as a longitudinal monitoring tool presents a significant challenge in analyzing and annotating seizures due to the vast amounts of collected data. Clinicians are not typically trained to identify seizures in non-standard EEG montages. Additionally, it is not feasible to manually annotate thousands of hours of data from multiple patients. It is of high interest to develop automated methods for detecting seizures based on the recorded data. Significant work has been done on automated machine learning (ML) frameworks to detect seizures based on changes in many types of recording modalities, such as scalp-EEG \cite{MLreview, DLreview}, ECG \cite{ECGdetect} and EMG \cite{EMGdetect}. However, research on wearable EEG is mainly unexplored and only initial findings have been published \cite{multi_2, bteabs1, bteabs2, bteabs3}. The development of ML algorithms for detecting seizures in wearable data involves significant adaptations to the pipelines due to differences in the properties and morphology of the data compared to standard modalities (vEEG). The use of bte-EEG for detecting focal seizures in practice is limited by the current automated frameworks' performance \cite{btefocal}. Despite the higher sensitivities achieved, when compared to seizure diaries, the number of false alarms is relatively high. The main ingredient of such automated methods is data. In literature, the majority of public datasets recorded from patients with epilepsy contain only full-scalp EEG data \cite{EEGdatasets}. There is a limited amount of open EEG datasets, and, to our knowledge, there are no publicly available datasets with recordings containing data from wearable devices from patients with epilepsy. The many published studies on ML for seizure detection are trained and tested on a heterogeneous cohort and use various validation methods. The training and test data are measured with different equipment, pruned in diverse ways and might contain non-continuous measurements, impeding generalization capabilities and decreasing the robustness of the seizure detection algorithms. The validation methods can vary depending on how the data is cleaned and the reported metrics are not standardized, creating a need for sharing data used in the development of such frameworks and a common evaluation pipeline \cite{szcore}. 

The dataset presented in this work was created with the objective of promoting the development of automated focal seizure detection frameworks in continuous wearable data. To the best of our knowledge, this is the first and largest phase 3 clinical study containing public multimodal (bte-EEG, ECG, EMG, ACC and GYR) wearable data recorded from patients with focal epilepsy.

\section*{Methods}

\subsection*{Study Design and Participants}

The SeizeIT2 project (clinicaltrials.gov: NCT04284072), a multicenter, prospective study, was carried out to validate the Sensor Dot device in adult and pediatric patients with epilepsy. Participants were included if they had a history of refractory epilepsy and were admitted to the Epilepsy Monitoring Unit (EMU) for long-term vEEG monitoring as a presurgical evaluation procedure. The exclusion criteria included patients with skin conditions or allergies that prevented the placement of the electrodes and adhesives or had implanted devices, such as neurostimulators or pacemakers. All participants provided written informed consent. The data collection started on January 10, 2020, and ended on June 30, 2022. The study was approved by the UZ Leuven ethics committee (approval ID: S63631, ClinicalTrials.gov, NCT04284072), anonymization and sharing of the data was also approved by the same committee (S67350 - amendment 1).

\begin{figure*}[h]
\centering
\includegraphics[width=0.6\linewidth]{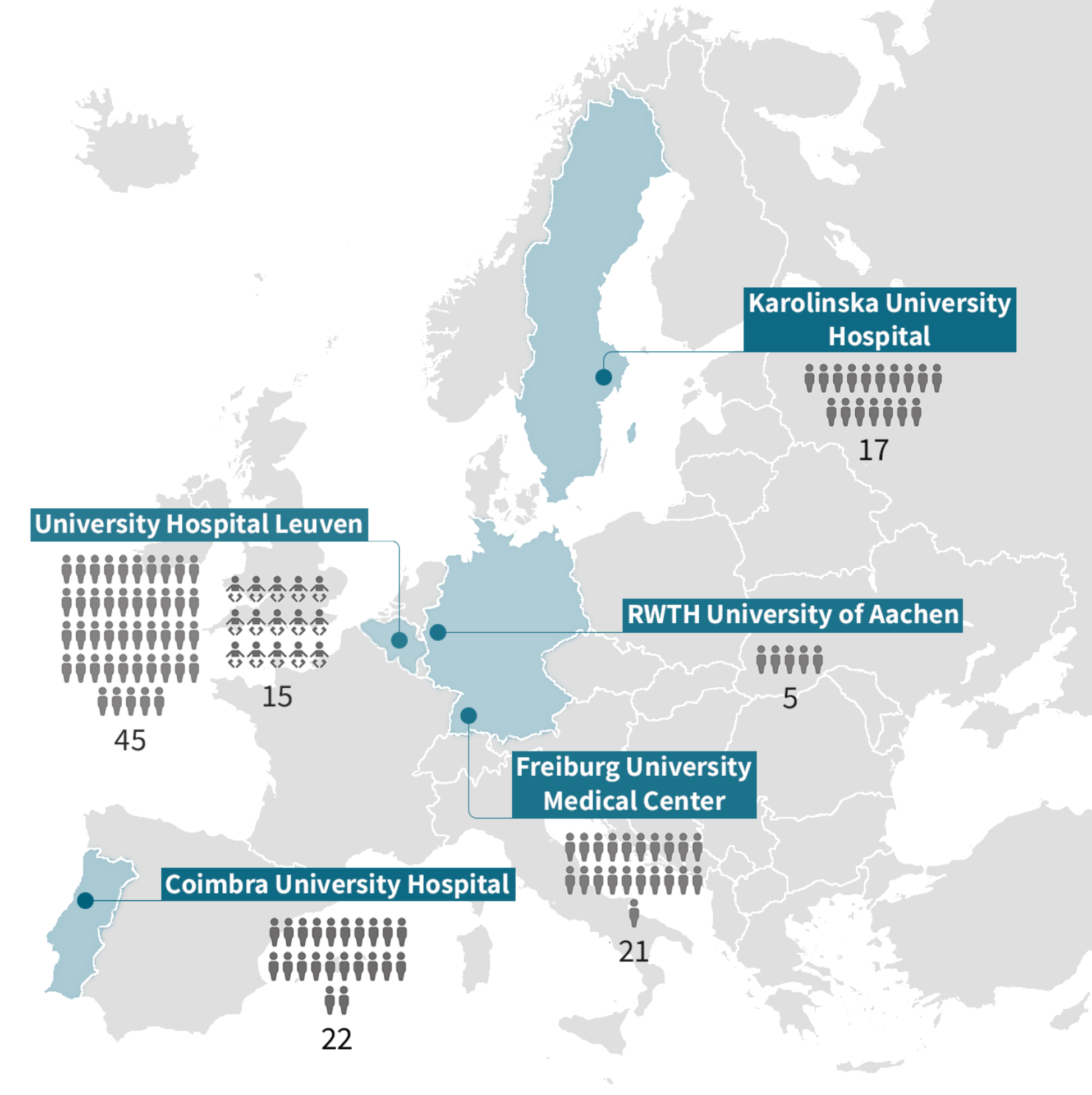}
\caption{Number of participants per EMU included in the SeizeIT2. The dataset includes patients from 5 different centers across 4 different European countries.}
\label{map}
\end{figure*}

The dataset comprises 125 patients (51 female, 41\%) from 5 different European EMUs: University Hospital Leuven (Belgium), Freiburg University Medical Center (Germany), RWTH University of Aachen (Germany), Karolinska University Hospital (Sweden) and Coimbra University Hospital (Portugal). Figure \ref{map} shows the distribution of the number of patients recorded in each center. The University Hospital Leuven was the only center that enrolled pediatric patients. The dataset includes only data from patients with focal epilepsy who experienced one or more seizure episodes during the monitoring period.

\subsection*{Recording Setup}

\begin{figure*}[h]
\centering
\includegraphics[width=0.8\linewidth]{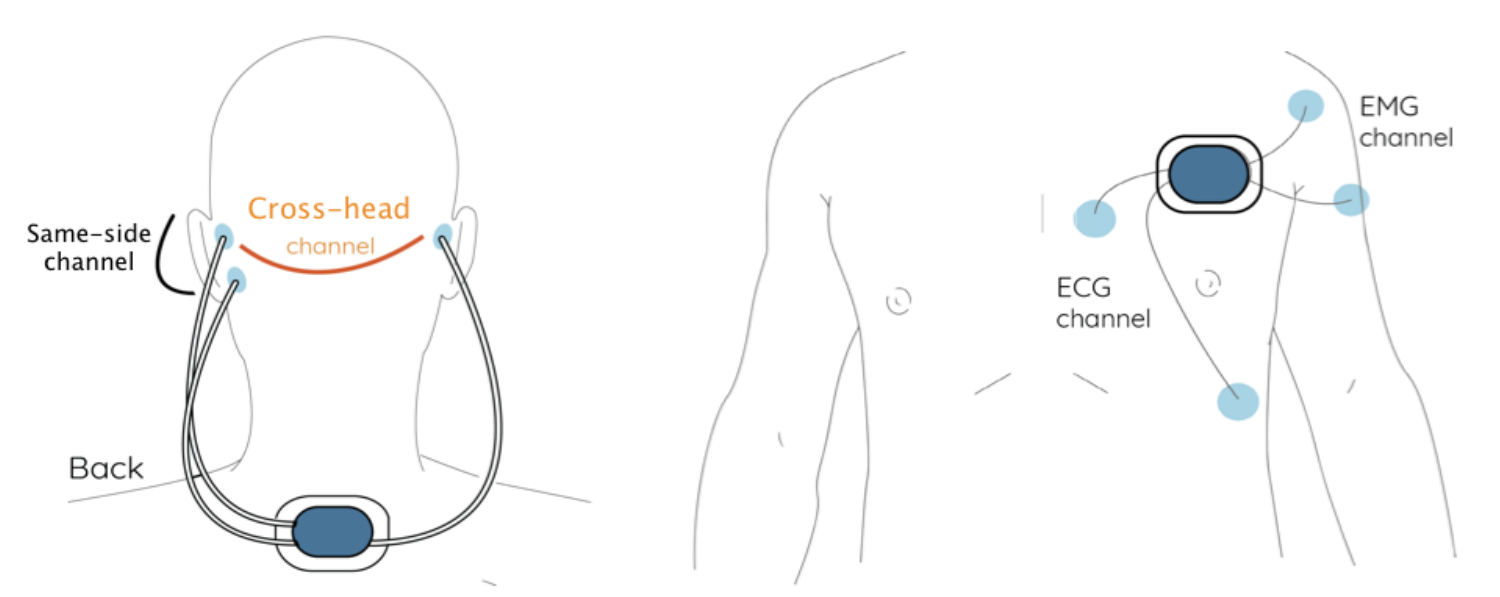}
\caption{Schematic of the placement of the Sensor Dot device. The example in the figure shows the setup for measuring patients with a suspected history of focal seizure occurrences with onset on the left hemisphere, with two electrodes behind the left ear and one on the right. Adapted from \cite{challenge}.}
\label{SD_setup}
\end{figure*}

The participants were recorded with the specific center's vEEG monitoring equipment, where the EEG electrodes were placed according to the 10-20 system or the 25-electrode array of the International Federation of Clinical Neurophysiology. The SD device was used to record wearable data simultaneously with the vEEG. The device has a size of 24.5 x 33.5 x 7.73 mm and weighs approximately 6.3 grams. The wearable device measures data at a sampling frequency of 250 Hz and has a battery life of approximately 24 hours. Two recording devices were used: one placed in the patient's upper back using a patch and connected to electrodes attached behind the ear, on the mastoid bone (EEG SD); another placed on the left side of the chest, with two electrodes extended to the lower left rib cage and the fourth intercostal space in the left parasternal position to measure ECG, and two electrodes extended to the left deltoid muscle to measure EMG data (ECG/EMG SD). The module itself contains accelerometers (ACC) and gyroscopes (GYR), which measured movement data at a sampling rate of 25 Hz. The SD setup is presented in Figure \ref{SD_setup}. The EEG SD electrode placement depended on the patient's medical history and is based on the seizure type and onset. When the seizures were suspected to originate from the left hemisphere, two electrodes were placed on the left side and one on the right side, forming one left same-side channel and one cross-head channel. Analogously, if seizures were suspected to originate from the right hemisphere, the same-side channel was derived from two electrodes placed behind the right ear. The dataset includes patients who were suspected to have generalized seizures (but had focal seizures) as well, and in this case, the cross-head channel was non-existent and replaced by an additional lateral channel by using two electrodes on each ear. The placement and impedance of each module were checked at the beginning and routinely during the monitoring sessions.

\subsection*{Dataset Content}

The complete dataset contains around 11 640 hours of wearable data. Four different modalities were recorded for most participants: bte-EEG, ECG, EMG and movement data. All participants' data within the dataset contain wearable bte-EEG. In 3\% of the dataset, ECG, EMG and movement data were not included due to technical failures or errors in the setup. In total, 886 focal seizures were recorded with the wearable device. The mean duration of the recorded seizures was 58 seconds, ranging between 3 seconds and 16 minutes. The majority of the seizures were focal aware (FA) and focal impaired awareness (FIA), with 317 and 393 occurrences, respectively. From the remaining seizures, 55 were focal-to-bilateral tonic clinic (FBTC), 12 were focal with unclear awareness status, 2 were subclinical focal seizures and 93 had unknown or unreported onset. There was a predominance of seizures with onset on the left hemisphere (44\%). In 12\% of the seizures, the onset was located in the right hemisphere, 1\% had a bilateral onset and in 43\% of the seizures the onset was unclear. Regarding localization, the seizure onsets were distributed over the central, frontal, temporal, occipital, parietal and insula lobes, with a predominance of temporal lobe seizures (30\%). Several seizures recorded could not be paired with a clear onset lobe (26\%). Table \ref{seiz_counts} shows detailed numbers of seizure occurrences and their respective lateralization and localization characteristics.

\begin{table*}[]
\centering
\caption{Number of seizures in the dataset for each type, lateralization and localization.}
\begin{tabular}{clcccccc}
\cline{3-8}
\multicolumn{1}{l}{}                                          &                                                & \textbf{FA} & \textbf{FIA} & \textbf{FBTC} & \textbf{Focal} & \textbf{Subclinical} & \textbf{Unknown} \\ \cline{2-8} 
\multicolumn{1}{l}{}                                          & \multicolumn{1}{l|}{\textbf{Total}}            & 317         & 393          & 55            & 98             & 17                   & 6                \\ \hline
\multicolumn{1}{c|}{\multirow{4}{*}{\textbf{Lateralization}}} & \multicolumn{1}{l|}{Left}                      & 170         & 161          & 26            & 20             & 10                   & -                \\
\multicolumn{1}{c|}{}                                         & \multicolumn{1}{l|}{Right}                     & 27          & 60           & 11            & 14             & 1                    & -                \\
\multicolumn{1}{c|}{}                                         & \multicolumn{1}{l|}{Bilateral}                 & -           & 5            & 2             & 2              & -                    & -                \\
\multicolumn{1}{c|}{}                                         & \multicolumn{1}{l|}{Unknown}                   & 120         & 167          & 16            & 62             & 6                    & 6                \\ \hline
\multicolumn{1}{c|}{\multirow{19}{*}{\textbf{Localization}}}  & \multicolumn{1}{l|}{Frontal}                   & 64          & 82           & 5             & 9              & -                    & -                \\
\multicolumn{1}{c|}{}                                         & \multicolumn{1}{l|}{Fronto-temporal}           & 81          & 11           & 8             & 6              & 1                    & -                \\
\multicolumn{1}{c|}{}                                         & \multicolumn{1}{l|}{Fronto-central}            & -           & -            & 2             & 6              & 2                    & -                \\
\multicolumn{1}{c|}{}                                         & \multicolumn{1}{l|}{Fronto-centro-temporal}    & 3           & 1            & -             & 2              & -                    & -                \\
\multicolumn{1}{c|}{}                                         & \multicolumn{1}{l|}{Fronto-centro-parietal}    & 12          & 1            & -             & 1              & -                    & -                \\
\multicolumn{1}{c|}{}                                         & \multicolumn{1}{l|}{Temporal}                  & 60          & 176          & 17            & 6              & 5                    & -                \\
\multicolumn{1}{c|}{}                                         & \multicolumn{1}{l|}{Temporo-occipital}         & -           & 1            & -             & 4              & -                    & -                \\
\multicolumn{1}{c|}{}                                         & \multicolumn{1}{l|}{Temporo-parietal}          & -           & 3            & 1             & -              & -                    & -                \\
\multicolumn{1}{c|}{}                                         & \multicolumn{1}{l|}{Temporo-parieto-occipital} & -           & -            & -             & 7              & -                    & -                \\
\multicolumn{1}{c|}{}                                         & \multicolumn{1}{l|}{Central}                   & 11          & -            & -             & -              & -                    & -                \\
\multicolumn{1}{c|}{}                                         & \multicolumn{1}{l|}{Centro-parietal}           & 6           & 10           & 1             & -              & -                    & -                \\
\multicolumn{1}{c|}{}                                         & \multicolumn{1}{l|}{Centro-temporal}           & 1           & -            & -             & -              & -                    & -                \\
\multicolumn{1}{c|}{}                                         & \multicolumn{1}{l|}{Centro-temporo-parietal}   & 6           & -            & -             & 1              & -                    & -                \\
\multicolumn{1}{c|}{}                                         & \multicolumn{1}{l|}{Occipital}                 & 5           & 6            & -             & 3              & 3                    & -                \\
\multicolumn{1}{c|}{}                                         & \multicolumn{1}{l|}{Occipito-parietal}         & -           & 1            & -             & -              & -                    & -                \\
\multicolumn{1}{c|}{}                                         & \multicolumn{1}{l|}{Parietal}                  & -           & -            & 1             & -              & -                    & -                \\
\multicolumn{1}{c|}{}                                         & \multicolumn{1}{l|}{Parieto-temporal}          & 2           & -            & 1             & -              & -                    & -                \\
\multicolumn{1}{c|}{}                                         & \multicolumn{1}{l|}{Insula}                    & 13          & 1            & 1             & 4              & -                    & -                \\
\multicolumn{1}{c|}{}                                         & \multicolumn{1}{l|}{Unknown}                   & 53          & 100          & 18            & 49             & 6                    & 6                \\ \hline
\end{tabular}
\label{seiz_counts}
\end{table*}

\section*{Data Records}

\begin{figure*}[h]
\centering
\resizebox{8 cm}{!}{%
\begin{tikzpicture}[%
  grow via three points={one child at (0.5,-0.7) and
  two children at (0.5,-0.7) and (0.5,-1.4)},
  edge from parent path={(\tikzparentnode.south) |- (\tikzchildnode.west)}
  ][transform canvas={scale=0.5}]
  \node {sz2}
    child { node [file] {\faFile \vspace{0.2cm} dataset\_description.json}}
    child { node [file] {\faFile \vspace{0.2cm} events.json}}
    child { node [file] {\faFile \vspace{0.2cm} participants.json}}
    child { node [file] {\faFile*[regular] \vspace{0.2cm} participants.tsv}}
    child { node {sub-001}}		
    child { node {sub-002}}
    child { node [selected] {sub-003}
        child { node {ses-001}
            child {node {eeg}
                child {node {\faFile[regular] \vspace{0.2cm} sub-001\_ses-01\_task-szMonitoring\_run-01\_eeg.edf \vspace{0.1cm}}}
                child {node {\faFile \vspace{0.2cm} sub-001\_ses-01\_task-szMonitoring\_run-01\_eeg.json \vspace{0.1cm}}}
                child {node {\faFile*[regular] \vspace{0.2cm} sub-001\_ses-01\_task-szMonitoring\_run-01\_events.tsv \vspace{0.1cm}}}
            }
            child [missing] {}
            child [missing] {}
            child [missing] {}
            child {node {ecg}
                child {node {\faFile[regular] \vspace{0.2cm} sub-001\_ses-01\_task-szMonitoring\_run-01\_ecg.edf \vspace{0.1cm}}}
                child {node {\faFile \vspace{0.2cm} sub-001\_ses-01\_task-szMonitoring\_run-01\_ecg.json \vspace{0.1cm}}}
            }
            child [missing] {}
            child [missing] {}
            child {node {emg}
                child {node {\faFile[regular] \vspace{0.2cm} sub-001\_ses-01\_task-szMonitoring\_run-01\_emg.edf \vspace{0.1cm}}}
                child {node {\faFile \vspace{0.2cm} sub-001\_ses-01\_task-szMonitoring\_run-01\_emg.json \vspace{0.1cm}}}
            }
            child [missing] {}
            child [missing] {}
            child {node {mov}
                child {node {\faFile[regular] \vspace{0.2cm} sub-001\_ses-01\_task-szMonitoring\_run-01\_mov.edf \vspace{0.1cm}}}
                child {node {\faFile \vspace{0.2cm} sub-001\_ses-01\_task-szMonitoring\_run-01\_mov.json \vspace{0.1cm}}}
            }
        }
        child [missing] {}
        child [missing] {}
        child [missing] {}
        child [missing] {}
        child [missing] {}
        child [missing] {}
        child [missing] {}
        child [missing] {}
        child [missing] {}
        child [missing] {}
        child [missing] {}
        child [missing] {}
        child [missing] {}
        }
    child [missing] {}
    child [missing] {}
    child [missing] {}
    child [missing] {}
    child [missing] {}
    child [missing] {}
    child [missing] {}
    child [missing] {}
    child [missing] {}
    child [missing] {}
    child [missing] {}
    child [missing] {}
    child [missing] {}
    child [missing] {}
    child [missing] {}
    child {node [file]{...}};
\end{tikzpicture}
}
\caption{File hierarchy of the repository.}
\label{file_hier}
\end{figure*}
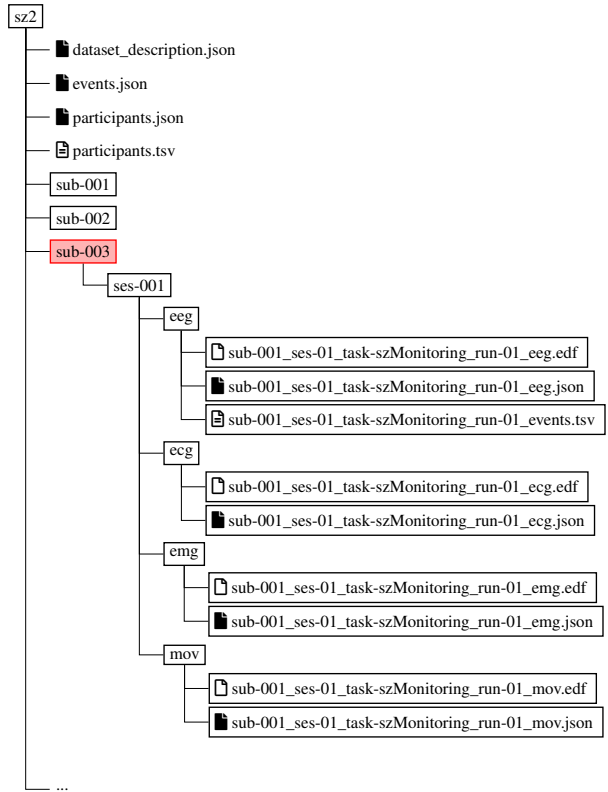

The dataset can be accessed at the OpenNeuro repository (https://openneuro.org/datasets/ds005873) \cite{repo} and conforms to the BIDS format \cite{bids}. The data structure in the repository is represented in Figure \ref{file_hier}.  The file 'dataset\_description.json' contains general information regarding the BIDS version used, licensing, authorship and acknowledgements. The 'events.json' file gathers categories and respective descriptions of all events annotated in the dataset, related to every event file associated with each recording. The information about the participants' sex is stored in the 'participants.tsv' file, with the associated 'participants.json' file with a description of the sex categories. Within the dataset, each participant's data is gathered in a separate folder ('sub-xxx'). Every main folder contains one subfolder ('ses-001') and within the latter, four different subfolders are organized to store each modality's data ('eeg', 'ecg', 'emg' and 'mov', for bte-EEG, ECG, EMG and movement- ACC and GYR- data respectively). For every modality's folder, each recording contains a '.edf' and a '.json' file, named with an extension '(...)\_run-XX\_' as an identifier. In the 'eeg' folder, there is an additional '.tsv' file associated to each recording with the extension '\_events'. The '.edf' files contain the data in European Data Format (EDF). The '.json' files include information about the sampling frequency, channel counts and placement, duration of the recordings and description of the task. The '\_events.tsv' files gather the annotations of each recording, with the fields described in the 'events.json' file: start time and duration of the recording, onset of the event, duration, type of event (in the case of a seizure event, the lateralization, localization and vigilance are included as well), confidence in the annotation and in which channels the event is visible.

\section*{Technical Validation}

\subsection*{Data quality}
All participants were monitored during the recordings, with daily checks on the wearable device placement and its impedance. Epileptologists and clinical neurophysiologists from each center evaluated the vEEG data of every participant to annotate seizure occurrences. The annotations were based on video records of the patients during monitoring and the full-scalp EEG data recorded with the EMU equipment. The annotations and the wearable data of this dataset were carefully aligned with the full-scalp EEG data. The seizures recorded, even if not visible in the data recorded with the SD device, are true seizures experienced by the patients. During data alignment, files in which the wearable data was completely corrupted (the recording contained only flat lines), were removed from the dataset.

\subsection*{Seizure detection}
\begin{figure*}[h]
\centering
\includegraphics[width=0.8\linewidth]{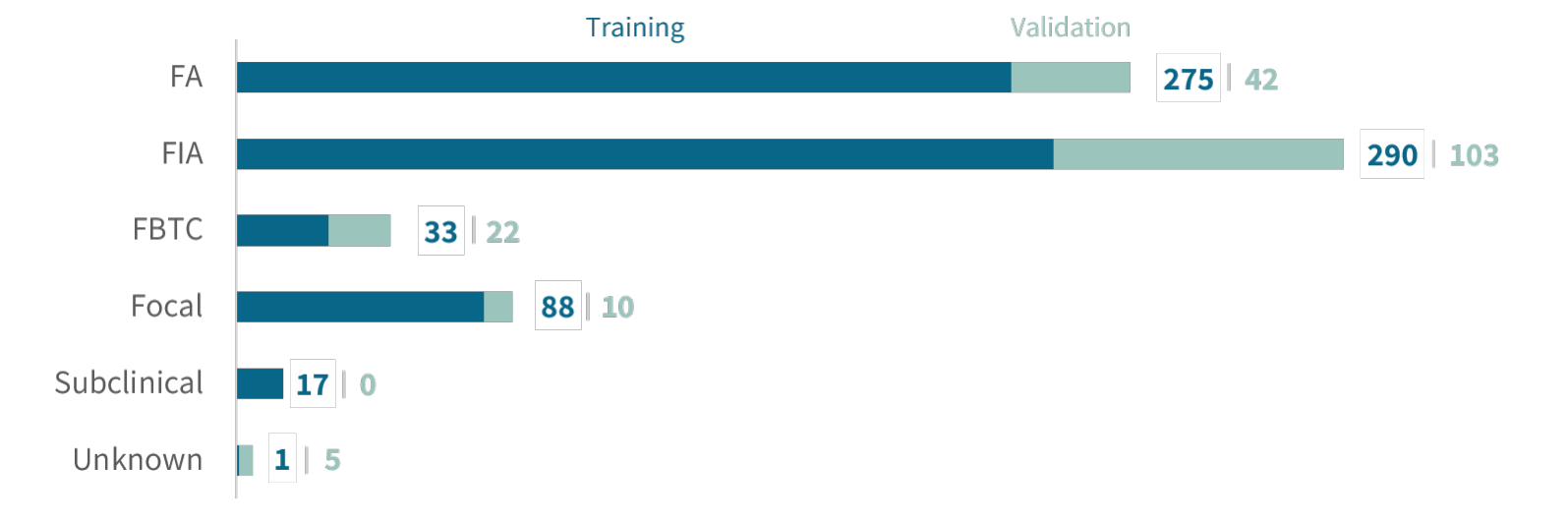}
\caption{Number of seizures per main type in the training and validation sets.}
\label{train_val_seiz}
\end{figure*}

The main objective of the dataset is to promote the development of automated seizure detection frameworks based on wearable data for continuous patient monitoring outside of the EMU. In this work, we implemented two different seizure detection methodologies, one based on a feature-based ML architecture and another using a deep learning framework. These serve as baseline methods for future seizure detection works with the dataset. In order to standardize the comparison between methods, we proposed a division between training and validation data. The training set contains 80\% of the data and the validation set 20\%. The division was made with efforts to keep the proportions of the number of seizures equal to the proportion of the amount of data in each set, as well as the number of patients from each center. Additionally, we attempted to keep the proportion of each seizure type equal within sets. The training set corresponds to the data from patients sub-001 to sub-096 and the validation set from patients sub-097 to sub-125 from the repository. The patient numbering is unrelated to any specific order. In total, 704 seizures were included in the training set and 182 in the validation set. The graph of Figure \ref{train_val_seiz} depicts the distribution of seizure types per set.

The feature-based method was based on a previous study \cite{kaat1}. In this work, a Support Vector Machine (SVM) model was used to detect seizures in bte-EEG recorded with hospital equipment. This method involves an initial pre-processing of the data with standard EEG filtering (1-25 Hz band-pass Butterworth filter) and data segmentation, creating 2-second EEG windows with a 50\% overlap used as input to the model. Segments were discarded if their root mean square amplitude was higher than 150 $\mu V$ or lower than 13 $\mu V$. A major difference between the work of Vandecasteele et al \cite{kaat1} and this manuscript is the number of channels. The previously used model was developed on 3-channel bte-EEG (two same-side channels, left and right, and one cross-head channel). The method was adapted to receive as input 2-channel bte-EEG, reducing the number of features to 42, since the same-side power asymmetry features are not relevant with this setup. Another difference is the label pruning. Previously, only the seizure data that was clearly visible on the wearable modality was included. With the SeizeIT2 data, all segments annotated as 'seizure' were included in the model's training. In order to balance the two classes, the majority class (background EEG) was undersampled to match the number of 'seizure' segments multiplied by a factor of five.

The second method is based on a DL architecture, the ChronoNet, that combines both convolutional and recurrent layers \cite{chrononet}. The architecture was developed initially for abnormal EEG classification. More recently, it was adapted for seizure detection on bte-EEG \cite{challenge}. Similar to the previous method, the input to the model is 2-second bte-EEG segments with 75\% overlap for the seizure data and 50\% for the background EEG. The pre-processing includes resampling the data to 250 Hz and three Butterworth filters (0.5 Hz high-pass, 60 Hz low-pass and 50 Hz notch filters). The training data is balanced in the same way as the feature-based method, choosing randomly five times the number of seizure segments for the background data segments.

\begin{figure*}[h]
\centering
\includegraphics[width=0.6\linewidth]{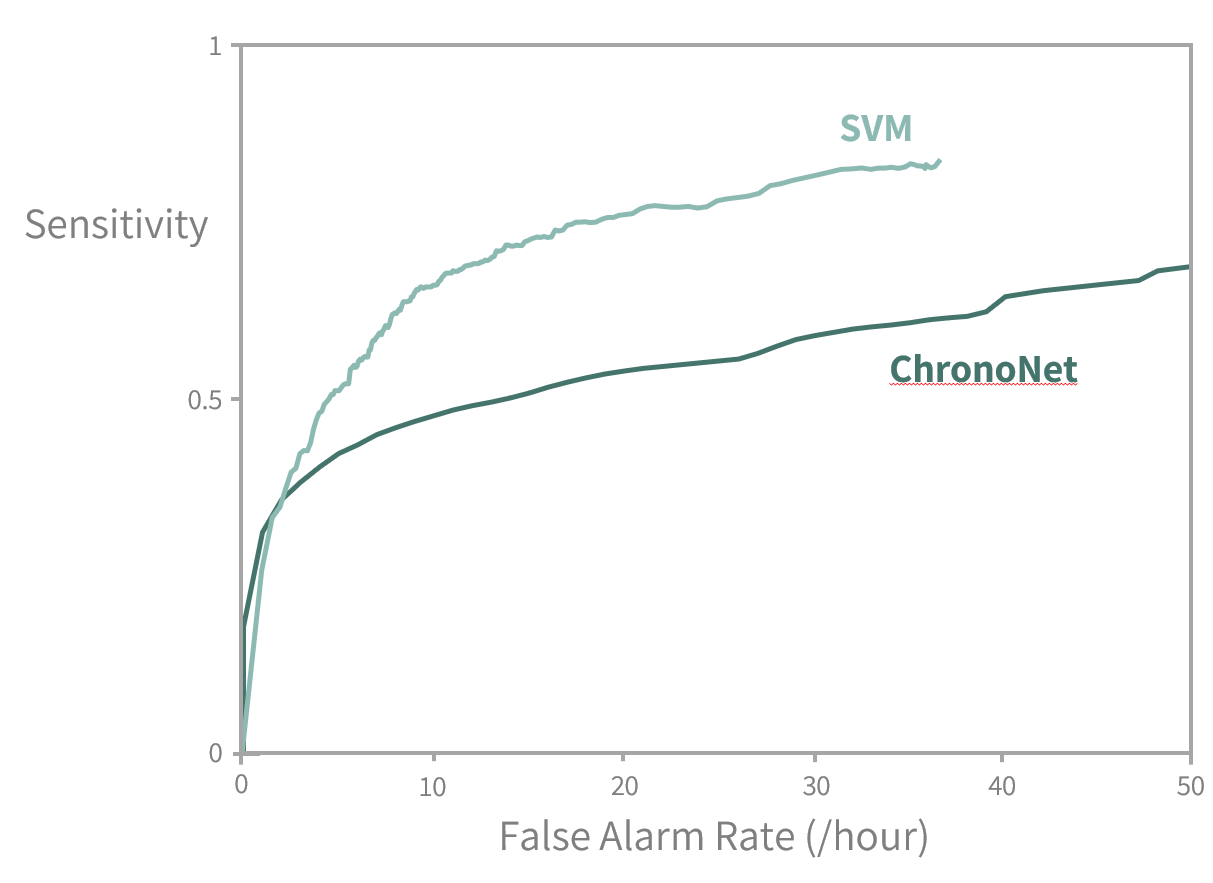}
\caption{Sensitivity - false alarm rate per hour curve}
\label{sens_fah}
\end{figure*}

The evaluation was done with both the traditional epoch-based and the any-overlap methods \cite{nedc}. Before evaluation, the model's classification probabilities went through a post-processing procedure. Firstly the segments with a root mean square amplitude below 13 $\mu V$ and above 150 $\mu V$ are discarded as potential seizure alarms. Furthermore, a positive seizure alarm was kept if, within a window of 10 seconds, there were at least 8 segments of 1 second classified as seizure. The sensitivity and the false alarm rate per hour were used to report the performance of the models. In this work, the area under the receiver operating characteristic curve (AUROC), the area under the precision-recall curve (AUPR) and the area under the sensitivity-normalized false alarm rate curve (AUSF) are also used as comparative metrics. The AUROC and AUPR were derived from metrics calculated using the traditional epoch-based method for computing sensitivity/recall, specificity and precision. The metrics values are presented in Table \ref{metrics}, as well as the sensitivity and the false-alarm rate calculated using a prediction threshold of 0.5. The sensitivity-false alarm rate per hour curve is presented in Figure \ref{sens_fah}. The figure displays the sensitivity values and correspondent false alarm rate per hour with varying threshold values for the predicted probabilities. Despite the ChronoNet method surpassing the SVM in all metrics associated to the area under the curves, the maximum sensitivity is lower and the trade-off between sensitivity and false alarm rate of the SVM is more suitable. In a clinical scenario, considering the use of automated models as an assistant tool for annotators, prioritizing sensitivity is more significant, since capturing all seizures leads to a better clinical evaluation of the patients. However, having a significantly high false alarm rate is not desirable.

\begin{table*}[h]
\centering
\caption{Performance of the SVM and ChronoNet models on the validation set.}
\label{metrics}
\begin{tabular}{llllll}
\cline{2-6}
                   & \textbf{Sens} & \textbf{FA/h} & \textbf{AUROC} & \textbf{AUPR} & \textbf{AUSF} \\ \hline
\textbf{SVM}       & 71.1          & 11.0         & 0.6995         & 0.0810        & 0.7229        \\
\textbf{ChronoNet} & 84.2          & 100.5        & 0.7921         & 0.1042        & 0.7953        \\ \hline
\end{tabular}
\end{table*}

The performance of the models presented in this study is still sub par to other methods published in the literature \cite{kaat1, multi_2}. The methods used for seizure detection were developed and validated in the initial iteration of the SeizeIT project, where hospital equipment was used to record patients. The wearable device introduces additional challenges for automated seizure detection frameworks such as added undesirable noise, lower data quality and decreased measurement reliability since the SD can be prone to technical errors and the measurement setup is not standardized. The clinical validation of the SD device in focal patients showed promising results regarding the patients' acceptance rate and clinical usability \cite{btefocal}. One of the major drawbacks was the limited performance of the automated methods to detect seizures. The presented dataset allows further development of these methods to implement such frameworks in clinical practice, and ease the burden on patients, care-givers and clinicians.

\section*{Usage Notes}

The dataset can be loaded and manipulated using the pipelines shared in our git repository (https://github.com/biomedepi/seizeit2). We include a Python data loader developed in Python 3.10.4, using the pyEDFlib v0.1.38 and pandas v2.2.3 packages.

\section*{Code availability}

All codes used to reproduce the work presented in this manuscript, including data pre-processing, model training and validation, can be accessed at https://github.com/biomedepi/seizeit2. The models were implemented with TensorFlow v2.10.0.

\section*{Acknowledgements}

We would like to acknowledge all clinicians who were involved in gathering the data and all researchers, mainly Thomas Strypsteen, Maarten Vanmarcke and Anna Martens, who aided in the development and validation of the shared code.

This work was funded by the European Union under the H2020-OTHER-EIT-HEALTH program (19263); Bijzonder Onderzoeksfonds (BOF) KU Leuven: “Prevalence of Epilepsy and Sleep Disturbances in Alzheimer Disease” (C24/18/097); Strategic basic research grant by Research Foundation Flanders (FWO) (for M. Bhagubai—1SB5922N); Research Foundation Flanders (FWO) Research Project, “Deep, personalized epileptic seizure detection” (G0D8321N); the Flemish Government (AI Research Program); M. De Vos, M. Bhagubai and C. Chatzichristos are affiliated to Leuven.AI - KU Leuven institute for AI, B-3000, Leuven, Belgium.

\section*{Author contributions statement}

Miguel Bhagubai: Writing - original draft, Writing - review and editing, Data curation, Data analysis, Methodology.
Christos Chatzichristos: Writing - review and editing, Data analysis, Methodology
Lauren Swinnen: Data collection, Data curation, Methodology
Jaiver Macea: Data collection, Data curation, Methodology
Jingwei Zhang: Data curation.
Lieven Lagae: Writing - review and editing, Data collection.
Katrien Jansen: Writing - review and editing, Data collection.
Andreas Schulze-Bonhage: Writing - review and editing, Data collection.
Francisco Sales: Writing - review and editing, Data collection.
Benno Mahler: Writing - review and editing, Data collection.
Yvonne Weber: Writing - review and editing, Data collection.
Wim Van Paesschen: Conceptualization, Funding acquisition, Project administration, Data collection, Writing - review and editing.
Maarten De Vos: Conceptualization, Funding acquisition, Project administration, Data collection, Writing - review and editing.

\section*{Competing interests}

The authors declare no competing interests.

\end{document}